\documentclass[twocolumn,showkeys,aps,prl,showpacs]{revtex4-1}
\UseRawInputEncoding
\usepackage{graphicx}
\usepackage[CJKbookmarks,dvipdfm,colorlinks,linkcolor=blue,citecolor=blue]{hyperref}
\usepackage{amsmath, amssymb}
\usepackage{makecell}
\usepackage{multirow}
\usepackage{CJK}
\usepackage{mathrsfs}
\usepackage{bm}
\usepackage{amsmath}
\usepackage{dcolumn}
\usepackage{epstopdf}
\usepackage{dsfont}
\usepackage{amssymb}
\usepackage{tabularx}
\usepackage{array}
\usepackage{float}
\usepackage{color}
\usepackage{epstopdf}
\usepackage{mathrsfs}
\usepackage{extarrows}

\begin{document}

\title{Electric-field induced half-metal in monolayer CrSBr}

\author{Hao-Tian Guo$^{1}$, San-Dong Guo$^{1}$$^{*}$ and Yee Sin Ang$^{2}$}
\affiliation{$^1$School of Electronic Engineering, Xi'an University of Posts and Telecommunications, Xi'an 710121, China}
\affiliation{$^2$Science, Mathematics and Technology (SMT), Singapore University of Technology and Design (SUTD), 8 Somapah Road, Singapore 487372, Singapore}

\begin{abstract}
Two-dimensional (2D) half-metallic materials  are highly desirable for nanoscale spintronic applications.
Here,  we propose a new mechanism that can  achieve half-metallicity in 2D  ferromagnetic (FM) material with two-layer magnetic atoms by electric field tuning.  We use a concrete example of experimentally synthesized CrSBr  monolayer  to  illustrate our  proposal through the first-principle calculations. It is found that the  half-metal can be achieved in  CrSBr  within  appropriate  electric field range, and the corresponding  amplitude of  electric field intensity is available in experiment.
Janus  monolayer $\mathrm{Cr_2S_2BrI}$ is constructed, which possesses built-in electric field due to broken horizontal mirror symmetry. However, $\mathrm{Cr_2S_2BrI}$ without and with applied external electric field is always a FM semiconductor. A possible memory device is also proposed based on CrSBr monolayer. Our works will stimulate
the application of 2D FM CrSBr in future spintronic nanodevices.

\end{abstract}
\keywords{Half-metal, Monolayer, Electric field~~~~~~~~~~~~~~~~~~~~~~~~~~~~~~~~~~Email:sandongyuwang@163.com}

\maketitle

\section{Introduction}
Spintronics provides a promising solution for developing  information technology
with high speed and low energy consumption by using the spin degrees of freedom of
electrons\cite{k1}. The half-metallic ferromagnets, where  one spin channel is
conducting while  the other spin channel is insulating
or semiconducting, possess  100\% spin polarized carriers around the
Fermi level, which are highly desired for high-performance spintronic devices by
filtering  the current into a single spin channel for realizing pure spin generation, injection,
and transport\cite{k2}. The three-dimensional (3D) half-metals have been widely investigated both in theory and in experiment\cite{k3}.
In order to realize  higher integration density, two-dimensional (2D) half-metallic ferromagnets may be preferred.
Although a number of 2D magnets  have been obtained experimentally\cite{k4,k5,k6},  a 2D  ferromagnetic (FM) half-metal is still missing.

Although there are various theoretical proposals of 2D half-metals\cite{k7,k8,k9}, no experimental evidence to prove the existence of intrinsic 2D half-metallicity has been reported. Fortunately,  bilayer A-type antiferromagnetic (AFM) materials (intralayer FM ordering and
interlayer AFM ordering) can be used to produce  half-metallic properties  by applying the out-of-plane electric field\cite{k10,k11}.
The underlying mechanism is that  the applied electric field makes the electrostatic potential of one constituent layer rise and that of the other layer decrease, giving rise to a semiconductor-metal phase transition. This strategy can also be used to realize the control of spin polarization in 2D A-type AFM semiconductors\cite{k12}.  Besides, the ferrovalley metal (FVM) and valley gapless semiconductor (VGS) can also be achieved in valleytronic bilayer AFM systems by electric field tuning\cite{k13}.

\begin{figure}
  \includegraphics[width=6cm]{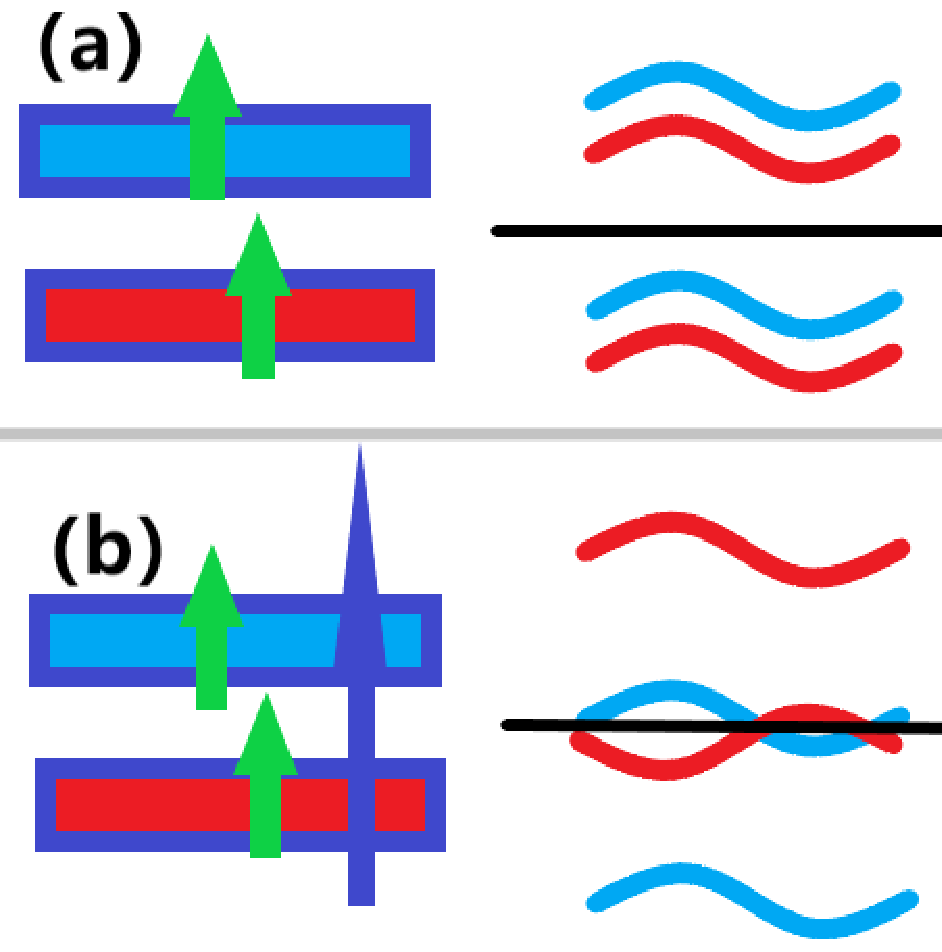}
  \caption{(Color online) (a): for a 2D material, the magnetic atoms have the same layer spin polarization, and the energy bands near the Fermi level are from the same spin channel, which have different layer character for both valence and conduction bands. (b): when an out-of-plane electric field is applied, the electronic bands in different layers will stagger, which can induce  half-metal. The green arrows mean spin, while the blue arrow shows an out-of-plane electric field.  The black horizontal lines mean Fermi level. }\label{sy}
\end{figure}

A natural question is whether this strategy or mechanism can be applied to 2D FM systems. Here,  we propose a new mechanism that can  achieve half-metallicity in 2D  FM material with two-layer magnetic atoms by electric field tuning (see \autoref{sy}).  For a 2D material, the two-layer magnetic atoms have the FM coupling, and the energy bands near the Fermi level are from the same spin channel with different layer character for both valence and conduction bands.  When an out-of-plane electric field is applied, the electronic bands in different layers will stagger, resulting in a half-metal.
To  illustrate our  proposal, a concrete example is  experimentally synthesized CrSBr  monolayer\cite{k14}, and both crystal structures and energy band structures  meet the above requirements\cite{k15,k16,k17}.
Calculated results show that the  half-metal   can indeed be achieved in  CrSBr  within  appropriate  electric field range, and the corresponding  amplitude of  electric field intensity is available in experiment ($\thicksim$0.4 $\mathrm{V/{\AA}}$\cite{fop3}). Our results pave the way for the
realization and application of 2D half-metal  in future spintronic
nanodevices.

\begin{figure}
  \includegraphics[width=7cm]{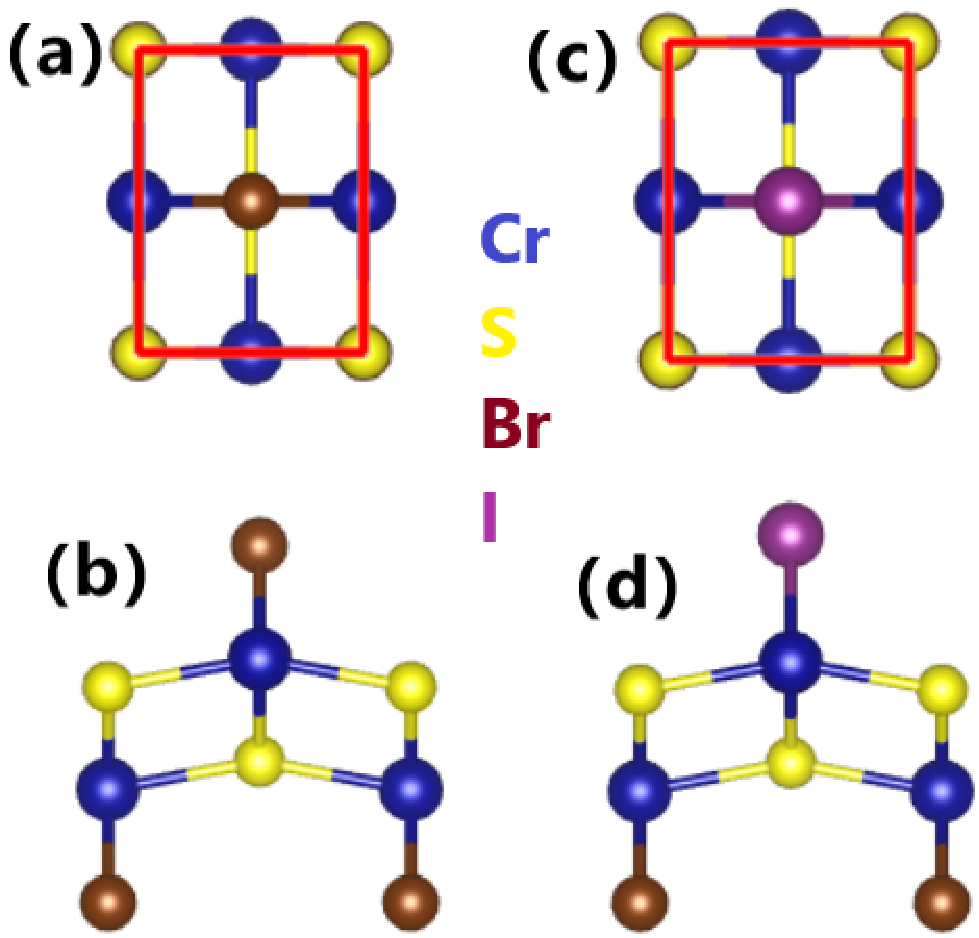}
  \caption{(Color online) For monolayer CrSBr (a,b) and Janus monolayer $\mathrm{Cr_2S_2BrI}$ (c,d):  top (a,c) and side (b,d) views of the  crystal structures.}\label{st}
\end{figure}

\section{Computational detail}
 Spin-polarized  first-principles calculations are carried out within density functional theory (DFT)\cite{1} by the projector augmented-wave (PAW) method, as implemented in the Vienna ab initio simulation package
(VASP)\cite{pv1,pv2,pv3}. We use the generalized gradient
approximation  of Perdew-Burke-Ernzerhof (PBE-GGA)\cite{pbe} as the exchange-correlation functional.
  The kinetic energy cutoff  of 500 eV,  total energy  convergence criterion of  $10^{-8}$ eV, and  force convergence criterion of 0.001 $\mathrm{eV.{\AA}^{-1}}$ are set to obtain accurate results.
To account for electron correlation of Cr-3$d$ orbitals, a Hubbard correction $U_{eff}$=3.00 eV\cite{k15,k16,k17}  is employed within the
rotationally invariant approach proposed by Dudarev et al.
To avoid out-of-plane interaction,  a vacuum of more than 16 $\mathrm{{\AA}}$ is adopted.
  We use a 18$\times$14$\times$1 Monkhorst-Pack k-point meshes to sample the Brillouin zone (BZ) for calculating electronic structures, elastic  and piezoelectric properties. The elastic stiffness tensor  $C_{ij}$  and piezoelectric stress tensor $e_{ij}$   are calculated by using strain-stress relationship (SSR) method and density functional perturbation theory (DFPT)\cite{pv6}, respectively. The  $C^{2D}_{ij}$/$e^{2D}_{ij}$ have been renormalized by   $C^{2D}_{ij}$=$L_z$$C^{3D}_{ij}$/$e^{2D}_{ij}$=$L_z$$e^{3D}_{ij}$, where the $L_z$ is  the length of unit cell along $z$ direction. The interatomic force constants (IFCs)  are obtained  by using  4$\times$4$\times$1 supercell within  finite displacement method, and  then calculate  phonon dispersion spectrum  by the  Phonopy code\cite{pv5}.

\section{main results}
Monolayer CrSBr possesses an orthorhombic crystal structure with a space group of $Pmmn$ (No.59)\cite{k14}, which has
 two distinct in-plane crystallographic directions (see \autoref{st} (a) and (b)). The CrSBr includes  six atomic layers in the sequence of Br-Cr-S-S-Cr-Br. The magnetic Cr atoms distribute in two layers, which reside in distorted octahedral coordination formed by S and Br atoms.
The lattice constants of CrSBr with FM ordering  are obtained by GGA+$U$, and the corresponding values $a$=3.584 $\mathrm{{\AA}}$ and $b$=4.825 $\mathrm{{\AA}}$.

\begin{figure*}
  \includegraphics[width=14cm]{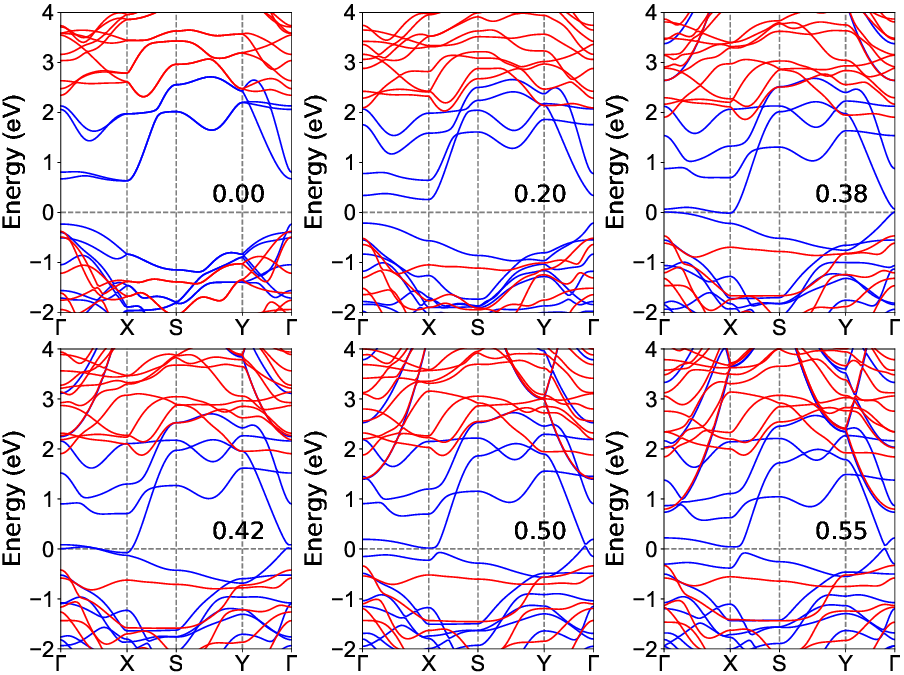}
\caption{(Color online) The energy  band structures of  CrSBr  at representative electric field $E$. The spin-up
and spin-down channels are depicted in blue and red.  An appropriate electric field intensity can induce  half-metal, for example $E$=0.42 $\mathrm{V/{\AA}}$. }\label{band}
\end{figure*}
\begin{figure}
  \includegraphics[width=8cm]{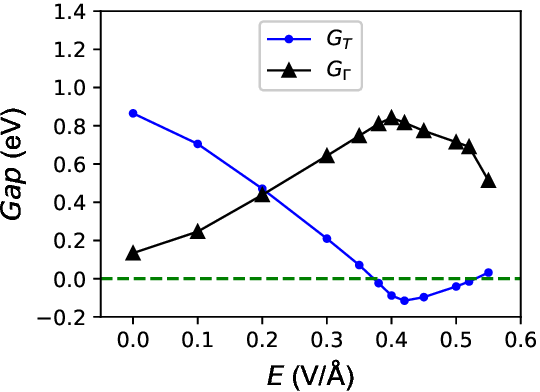}
  \caption{(Color online)For CrSBr, the  global gap ($G_{T}$) and the energy difference ($G_{\Gamma}$) at $\Gamma$ point for the first two conduction bands as a function of $E$. }\label{gap}
\end{figure}
Electric field is a very effective way to tune electronic structures, topological properties and  magnetic anisotropy energy of 2D materials\cite{k10,k11,k12,k13,e1,e2,e3}.
Here, the  electric field effects  on  magnetic properties and electric structures  in CrSBr are investigated.
 Firstly, we determine the magnetic ground state under the electric field, and four magnetic configurations (FIG.1 of  electronic supplementary information (ESI)) are considered, including FM case and three AFM cases (AFM, AFM-x and AFM-y).
 The energy differences per formula unit between AFM/AFM-x/AFM-y and FM as a function of electric field $E$ are shown in FIG.2 of ESI. For AFM-x case, the result converges to a non-magnetic state, which leads to very high energy. Calculated results show that the FM state is always  ground state within  considered $E$ range.

The energy band structures of  CrSBr under representative  $E$ are shown in \autoref{band}, and the  global gap ($G_{T}$)  and the energy difference ($G_{\Gamma}$) at $\Gamma$ point for the first two conduction bands as a function of $E$ are plotted in \autoref{gap}.
Without electric field, CrSBr is a FM semiconductor with gap value of 0.865 eV, and the valence band maximum (VBM) and conduction band bottom (CBM) are at  $\Gamma$ and  one point close to X, which are from the same spin-up channel. This provides possibility to achieve half-metal in CrSBr by electric field tuning. With increasing  $E$, the global gap decreases, and  a semiconductor-metal phase transition can be observed at about  $E$$=$0.38  $\mathrm{V/{\AA}}$. At about $E$$=$0.55  $\mathrm{V/{\AA}}$, a gap of  24 meV is produced, meaning a  metal-semiconductor phase  transition.
Calculated results show that $G_{\Gamma}$ firstly increases, and then decreases.
For  0.38  $\mathrm{V/{\AA}}$$<$$E$$<$0.55  $\mathrm{V/{\AA}}$,  2D half-metal can be realized in CrSBr.
Here, we define half-metallic gap $G_{HM}$ as the smaller
of $E_{cb}$ and $E_{vt}$, where $E_{cb}$/$E_{vt}$ is the bottom/top energy of the spin-down conduction/valence bands with respect to the Fermi (absolute value), which is  rough estimates for the minimal energy for spin flip excitation.
The $G_{HM}$ vs $E$ is plotted in FIG.3 of ESI, which are all larger than 0.33 eV.

This out-of-plane electric field creates a layer-dependent electrostatic potential\cite{k10,k11,k12,k13}, and the electronic bands in different Cr layers will stagger, which  gives rise to half-metallic property in CrSBr. To clearly see this, two Cr layer-characters energy band structures  are plotted in FIG.4 of ESI at $E$$=$0.35 $\mathrm{V/{\AA}}$ and 0.42 $\mathrm{V/{\AA}}$, which shows that  the energy band from down-layer Cr is shifted toward higher energy  with respect to one of up-layer.  Experimentally, the CrSBr monolayer has been achieved\cite{k14}. Recently, an intense electric field larger than 0.4 $\mathrm{V/{\AA}}$ has been realized  in 2D materials by dual ionic gating\cite{fop3}. These  provide  very good foundation to experimentally produce half-metallic property  in CrSBr.

 The out-of-plane built-in electric field  is equivalent to an  external electric field\cite{ar1}. The CrSBr possesses spatial centrosymmetry, producing no built-in electric field. Here, Janus monolayer $\mathrm{Cr_2S_2BrI}$ is constructed by  replacing one of two Br  layers with I atoms in monolayer  CrSBr. The schematic crystal structures of $\mathrm{Cr_2S_2BrI}$ are shown in \autoref{st} (c) and (d),  which has  $Pmm2$ space group (No.25).
 Due to broken horizontal mirror symmetry, an out-of-plane built-in electric field can be produced in $\mathrm{Cr_2S_2BrI}$.
The energy differences between AFM/AFM-x/AFM-y and FM configurations is 0.097/0.095/0.070 eV, and the positive values  confirm that $\mathrm{Cr_2S_2BrI}$ is FM ground state.   The optimized lattice constant $a$ and $b$  are 3.662 $\mathrm{{\AA}}$  and 4.816  $\mathrm{{\AA}}$ with FM ordering by GGA+$U$.

To  verify its dynamical stability, the phonon dispersions of  $\mathrm{Cr_2S_2BrI}$ is calculated,   as shown in  FIG.5 of ESI.  No obvious imaginary frequencies are observed in the whole BZ,  implying   that $\mathrm{Cr_2S_2BrI}$ is dynamically stable.
To further confirm its thermal stability, the  ab-initio molecular dynamics (AIMD) simulations are carried out
 with a 4$\times$3$\times$1 supercell and a time step
of 1 fs at 300 K for 8 ps.  According to FIG.6 of ESI,  the
energy  fluctuates  within a small range during the whole simulation time, and the snapshot  shows no structural transitions at the end of the AIMD simulations,  which manifest its thermal stability.
The linear elastic constants can be used to determine the  mechanical stability of  $\mathrm{Cr_2S_2BrI}$.
Due to  $Pmm2$ space group, there are  four  independent elastic constants: $C_{11}$=81.37 $\mathrm{Nm^{-1}}$, $C_{12}$=9.48 $\mathrm{Nm^{-1}}$,  $C_{22}$=98.57 $\mathrm{Nm^{-1}}$ and   $C_{66}$=21.36 $\mathrm{Nm^{-1}}$, which meet Born-Huang
criteria of  mechanical stability  ($C_{11}>0$, $C_{22}>0$, $C_{66}>0$ and  $C_{11}-C_{12}>0$),  thereby verifying its mechanical
stability.

\begin{figure}
  \includegraphics[width=8cm]{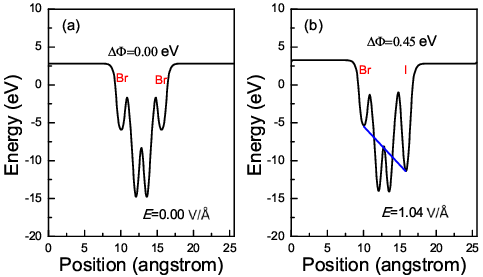}
  \caption{(Color online)For monolayer $\mathrm{CrSBr}$ (a) and  $\mathrm{Cr_2S_2BrI}$ (b), the planar averaged electrostatic potential energy variation along $z$ direction. $\Delta\Phi$ is the potential energy difference across the layer. $E$ means built-in electric field. }\label{pe}
\end{figure}

The planar average of the electrostatic potential energy of  CrSBr and $\mathrm{Cr_2S_2BrI}$  along $z$ direction are
shown in \autoref{pe}.  An electrostatic potential gradient ($\Delta \Phi$)  is
related to the work function change of the structure. Due to horizontal mirror symmetry, $\Delta \Phi$ and inherent electric field of  CrSBr are zero. Due to  mirror asymmetry of $\mathrm{Cr_2S_2BrI}$, there is an inherent electric field with the magnitude
of about  1.04 $\mathrm{V/{\AA}}$, and $\Delta \Phi$ of about 0.45 eV, which are due to the
electron redistribution caused by different  electronegativity  of Br and I atoms.
The magnitude of  inherent electric field for $\mathrm{Cr_2S_2BrI}$ is larger than an external electric field, which can make CrSBr become a half-metal.

\begin{figure}
  \includegraphics[width=8cm]{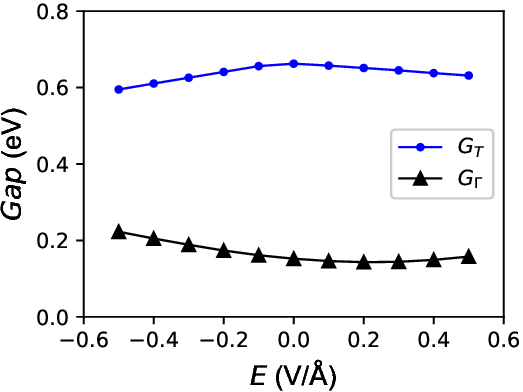}
  \caption{(Color online)For $\mathrm{Cr_2S_2BrI}$, the  global gap ($G_{T}$) and the energy difference ($G_{\Gamma}$) at $\Gamma$ point for the first two conduction bands as a function of $E$. }\label{gap-1}
\end{figure}

Next, the electronic structures of $\mathrm{Cr_2S_2BrI}$ along with electric field effects are investigated. Due to  mirror asymmetry of $\mathrm{Cr_2S_2BrI}$, the electric fields along the $+z$ and $-z$ directions are not equivalent.
The energy differences per formula unit between AFM/AFM-x/AFM-y and FM orderings as a function of electric field $E$ are shown in FIG.7 of ESI, which shows that the FM state is always  ground state of $\mathrm{Cr_2S_2BrI}$ within  considered $E$ range.
The energy band structures of  $\mathrm{Cr_2S_2BrI}$  under representative  $E$ are shown in FIG.8 and FIG.9 of ESI, and the  global gap ($G_{T}$)  and the energy difference ($G_{\Gamma}$) at $\Gamma$ point for the first two conduction bands as a function of $E$ are plotted in \autoref{gap-1}.
Without electric field, $\mathrm{Cr_2S_2BrI}$ is still a FM semiconductor with gap value of 0.663 eV, and the VBM and CBM are at  $\Gamma$ and  one point close to X, which are from the same spin-up channel. These are very similar to those of CrSBr.
Calculated results show that no semiconductor-metal phase transition is produced within considered $E$ range (-0.5 $\mathrm{V/{\AA}}$ to 0.5 $\mathrm{V/{\AA}}$).
It is found that electric filed has small effects on $G_{T}$ and $G_{\Gamma}$ of $\mathrm{Cr_2S_2BrI}$. Therefore, it is not possible to achieve  half-metal  by building Janus structures with built-in electric field.

However, compared with CrSBr, $\mathrm{Cr_2S_2BrI}$ possesses  piezoelectricity due to broken spatial inversion symmetry.
 By using  Voigt notation,  only considering the in-plane strain and stress\cite{q7}, the  piezoelectric stress   and strain tensors of $\mathrm{Cr_2S_2BrI}$ can be written as:
 \begin{equation}\label{pe1-1}
 e=\left(
    \begin{array}{ccc}
     0 & 0 & 0 \\
     0 & 0 & 0 \\
      e_{31} & e_{32} & 0 \\
    \end{array}
  \right)
    \end{equation}

  \begin{equation}\label{pe1-2}
  d= \left(
    \begin{array}{ccc}
      0 & 0 & 0 \\
       0 & 0 & 0 \\
      d_{31} & d_{32} &0 \\
    \end{array}
  \right)
\end{equation}
The $e_{31}$ and $e_{32}$ can be attained by DFPT, and the
$d_{31}$ and $d_{32}$ can be  derived by  $e_{ik}$=$d_{ij}C_{jk}$:
\begin{equation}\label{pe2}
     d_{31}=\frac{e_{31}C_{22}-e_{32}C_{12}}{C_{11}C_{22}-C_{12}^2}~and~
     d_{32}=\frac{e_{32}C_{11}-e_{31}C_{12}}{C_{11}C_{22}-C_{12}^2}
\end{equation}
The calculated $e_{31}$/$e_{32}$ is 0.708$\times$$10^{-10}$/0.016$\times$$10^{-10}$ C/m. And then, the  $d_{31}$/$d_{32}$ of $\mathrm{Cr_2S_2BrI}$ can be attained from \autoref{pe2}, and the corresponding value is 0.878/-0.068 pm/V, showing very strong in-plane anisotropy. The $d_{31}$  of   $\mathrm{Cr_2S_2BrI}$ is compared with or  higher  than ones of many known 2D  materials\cite{q7,q8,q9,q10,q11}.

\begin{figure}
  \includegraphics[width=7cm]{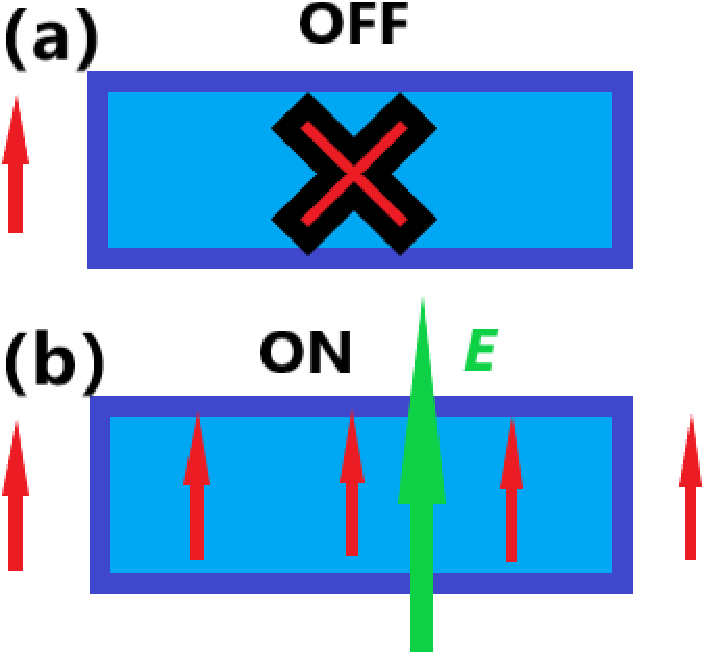}
  \caption{(Color online)Schematic of a possible spintronic device. The red arrows represent spin, while the green arrow represents an out-of-plane electric field.  }\label{qj}
\end{figure}

\section{discussion and Conclusion}
The  half-metallic property can be induced by applied out-of-plane  electric field in CrSBr,  which provides a possibility to realize  memory device (see \autoref{qj}). For example, two FM electrodes  are set in the same direction for  source and drain electrodes.  Without the out-of-plane electric
field, no electrons can propagate through the CrSBr due to  semiconducting property  (``off" state)) (\autoref{qj} (a)).
When an appropriate out-of-plane external electric field is applied,  spin-up electrons can propagate through the CrSBr because of half-metallic property (``on" state)) (\autoref{qj} (b)).  Here, data writing in the
memory is realized by turning  on and off out-of-plane external electric field,
and data reading is realized by detecting the electrical signals.

In summary,  a new mechanism is proposed to  achieve half-metallicity in 2D FM systems with two-layer magnetic atoms by electric field tuning.  A concrete example of experimentally synthesized CrSBr  monolayer  is used  to  illustrate our  proposal through the first-principle calculations. The  half-metal can indeed be achieved in  CrSBr  within  experimentally available  electric field range.
Janus  monolayer $\mathrm{Cr_2S_2BrI}$ is constructed with built-in electric field, but it can not be used to induce half-metallic property. In fact,  our work provides a feasible and general
approach to achieve half-metal.
The method, analysis and results  can be readily extended to  VSF, VSeF, VSeBr, CrSCl,  CrSI, CrSeBr and CrSeI\cite{tg}, which have the same structure and FM  configuration with CrSBr. Based on these  monolayers, an applied Vertical electric field can  realize half-metal.

~~~~\\
\textbf{SUPPLEMENTARY MATERIAL}
\\
See the supplementary material for magnetic configurations; the energy difference between AFM and FM ordering; half-metallic gap; the related energy band structures; the phonon dispersions and AIMD results.

~~~~\\
\textbf{Conflicts of interest}
\\
There are no conflicts to declare.

~~~~~\\

\begin{acknowledgments}
This work is supported by Natural Science Basis Research Plan in Shaanxi Province of China  (2021JM-456). We are grateful to Shanxi Supercomputing Center of China, and the calculations were performed on TianHe-2.
\end{acknowledgments}

\end{document}